\documentclass[usenatbib]{mn3e} 
\usepackage{amsmath,amssymb}
\usepackage[figuresleft]{rotating}     
\usepackage{tablefootnote}   
\usepackage{graphicx}         
\usepackage{gensymb}        
\usepackage{color}               
\usepackage{natbib}
\usepackage{times}
\usepackage{fixltx2e} 
\usepackage[flushleft]{threeparttable}
\usepackage{hyperref}     

\usepackage{url}
\usepackage{caption}        
\newcommand{\ramses}{{\small RAMSES}}

\newcommand{\cupid}{{\small CUPID} }
\newcommand{\cupidtext}{{\small CUPID}}

\newcommand{\HH}{{\rm H}$_2$ }

\newcommand{\Msol}{{\,{\rm M}}_\odot}

\newcommand{\kpc} {{\,\rm kpc}}
\newcommand{\pc} {{\,\rm pc}} 
\newcommand{\K} {{\,\rm K}} 
\newcommand{\cc}{{\,\rm {cm^{-3}}}}

\newcommand{\kmsec}{{\,\rm {km\,s^{-1}} }}

\def\Myr{\,{\rm Myr}}

\newcommand{\kms}{{\,\rm {km\,s^{-1}} }} 

\newcommand{\Msun}{M_\odot}

\newcommand{\rthres}{\rho_{\rm thres}}
\newcommand{\sthres}{\rho_{\rm saddle}}
\newcommand{\ap}{\alpha_{\rm vir}}
\newcommand{\pv}{P_{\rm vir,ext}}
\newcommand{\rg}{R_{\rm G}}

\newcommand{\bible}{\cite{Agertz:2013aa} }

\newcommand{\Sol}{\cite{Solomon:1987aa} }

\title[Properties of GMCs]{Physical properties and scaling relations of molecular clouds: the effect of stellar feedback}\author[Kearn Grisdale et al.] {\parbox[t]{\textwidth}{Kearn Grisdale$^{1,4}$\thanks{kearn.grisdale@physics.ox.ac.uk}, Oscar Agertz$^{2}$, Florent Renaud$^{2,4}$, and Alessandro B. Romeo$^{3}$}\vspace*{6pt}\\
  $^1$ Sub-department of Astrophysics, University of Oxford, Keble Road, Oxford OX1 3RH\\ 
  $^2$ Lund Observatory, Department of Astronomy and Theoretical Physics, Box 43, 221 00 Lund, Sweden\\
  $^3$ Department of Space, Earth and Environment, Chalmers University of Technology, SE-41296 Gothenburg, Sweden\\
  $^4$ Department of Physics, University of Surrey, Guildford GU2 7XH, United Kingdom\\
  }
\date{\today}
\begin{document}
\maketitle
\graphicspath{ {Figures/} }

\begin{abstract} 
Using hydrodynamical simulations of entire galactic discs similar to the Milky Way, reaching $4.6\pc$ resolution, we study the origins of observed physical properties of giant molecular clouds (GMCs). We find that efficient stellar feedback is a necessary ingredient in order to develop a realistic interstellar medium (ISM), leading to molecular cloud masses, sizes, velocity dispersions and virial parameters in excellent agreement with Milky Way observations. GMC scaling relations observed in the Milky Way, such as the mass-size ($M$--$R$), velocity dispersion-size ($\sigma$--$R$), and the $\sigma$--$R\Sigma$ relations, are reproduced in a feedback driven ISM when observed in projection, with $M\propto R^{2.3}$ and $\sigma\propto R^{0.56}$. When analysed in 3D, GMC scaling relations steepen significantly, indicating potential limitations of our understanding of molecular cloud 3D structure from observations. Furthermore, we demonstrate how a GMC population's underlying distribution of virial parameters can strongly influence the scatter in derived scaling relations. Finally, we show that GMCs with nearly identical global properties exist in different evolutionary stages, where a majority of clouds being either gravitationally bound or expanding, but with a significant fraction being compressed by external ISM pressure, at all times.

\end{abstract}


\section{Introduction}
\label{sect:intro}

The role of stellar feedback processes (henceforth feedback) such as supernovae explosions, stellar winds and ionising radiation from massive stars, in the evolution of galaxies is highly debated topic \citep{Dekel:1986aa,Efstathiou:2000aa,Hopkins:2014aa,Agertz:2016aa,Grisdale:2016aa}. It has been well established that stars form in Giant Molecular Clouds (GMCs) \citep[][and references within]{MacLow:2004aa,McKee:2007aa} and it is here that feedback will have its first impact and affect the local star formation process. 

The definition of GMCs, or at least of their external boundaries is somewhat arbitrary. Observationally, it often derives from the tracer of dense gas used \citep[e.g. CO, see][and references within]{Dobbs:2014aa}. Comparable definitions can be adopted in simulations, or be extended to larger volumes encompassing the bound mass of clouds, irrespective of the atomic or molecular nature of the cloud and its envelope, to better account for dynamical properties. In this paper, we adopt a criterion based on the density of the gas (and not its kinematics), to allow comparisons between observational data and simulations.

In the past four decades, significant work has been done on observing and quantifying properties of GMCs both in the Milky Way (MW) and other galaxies \citep[e.g.][]{Larson:1981aa,Solomon:1987aa,Heyer:2009aa,Roman-Duval:2010aa,Rice:2016aa,Miville-Deschenes:2017aa}, and clumpy galaxies at high redshift \citep[e.g.][]{Swinbank:2015aa}. These studies have highlighted that observed GMCs feature a wide range of properties. For example, \Sol used CO observations to measure sizes, velocity dispersion and masses of GMCs in the solar neighbourhood. Such surveys have been complemented by \cite{Heyer:2009aa}, henceforth H09, and extended to 8107 clouds over the entire Galactic disc by \cite{Miville-Deschenes:2017aa}, henceforth MD17, who found masses in the range $10\Msol\lesssim M\lesssim10^{7}\Msol$, radii $0.5\pc\lesssim R \lesssim 200\pc$ and velocity dispersions $0\kms<\sigma\lesssim10\kms$.
 
\cite{Rosolowsky:2005aa} and \cite{Rosolowsky:2007aa} \citep[see also][]{Rosolowsky:2003aa} found similar value for GMCs in M64 and M33 respectively. \cite{Hughes:2013aa} carried out a comparative study of the GMCs in M51, M33, and the Large Magellanic Cloud and highlighted that the different galactic environments (e.g. arms, inter-arms) produce different populations of GMCs, characterised by different distributions of surface density $\Sigma$, radius and velocity dispersion, possibly due to differences in ISM pressure \citep[see e.g.][]{Meidt:2013aa}.

In his pioneering work, \cite{Larson:1981aa} found correlations between the properties of GMCs, often referred to as ``Larson's scaling laws":
\begin{equation}
	\sigma \propto R^{a},
	\label{eq:larson1}
\end{equation}
\begin{equation}
	M \propto R^{b},
	\label{eq:larson2}
\end{equation}
\begin{equation}
	\sigma \propto (R\Sigma)^{c}.
	\label{eq:larson3}
\end{equation}
Canonical power-law indexes for these relations are $a\approx 0.5$ and $b\approx 2$ \citep{Larson:1981aa, Solomon:1987aa}. Their values changed with improved observations, with recent data on MW clouds putting them at $a = 0.63\pm0.30$, $b=2.2\pm 0.2$ and $c=0.43\pm 0.14$ \citep[][]{Falgarone:2009aa,Heyer:2009aa,Roman-Duval:2010aa,Miville-Deschenes:2017aa}. GMCs in other galaxies have been found to also follow the Larson relations \citep[][]{Hughes:2013aa,Faesi:2016aa,Tosaki:2017aa}.Larson's scaling laws not only characterize the physical state of GMCs, but also have an impact on the gravitational instability of galaxy discs \citep[e.g.][]{Elmegreen:1996aa,Romeo:2010aa,Agertz:2015aa} at scales smaller than, or comparable to, the disc scale height \citep{Hoffmann:2012aa,Romeo:2014aa}.
 
Understanding the origins of observed distributions of GMC properties and scaling relations have been an intense research topic in the past few decades \citep[e.g.][]{Ballesteros-Paredes:2011aa,Ballesteros-Paredes:2012aa,Meidt:2013aa,Kauffmann:2013aa}. One approach has been to simulate isolated regions (e.g. $\leq1\kpc^{3}$ boxes) of the ISM \citep[for example][to name a few]{Audit:2010aa,Walch:2015aa,Ibanez-Mejia:2016aa,Padoan:2016aa,Iffrig:2017aa}. Due to the limited size of the simulated volume, such simulations ISM simulations often reach sub-parsec resolution, and are thus able to resolve the internal structure of GMCs. Many of these models often lack (possibly important) physical processes, such as large scale galactic rotation/shear, self-gravity and models of star formation and stellar feedback. Despite such simplifications, modern work focusing a purely supernova driven ISM, without gravity \citep[e.g.][]{Padoan:2016aa} or with \citep[][]{Ibanez-Mejia:2016aa} produce GMC populations that compare favourably to observations.

The level of boundness of a clouds is another fundamental property, often defined via the \emph{virial parameter} \citep{Bertoldi:1992aa},
\begin{equation}
	\ap \equiv \frac{5\sigma^{2}R}{GM},
	\label{eq:virial}
\end{equation}
which is the ratio between twice the kinetic energy and the potential energy. Approximately half of observed GMCs tend to be in or near virial equilibrium ($\ap = 1$) or gravitationally bound ($\ap \lesssim 2$). A significant number of GMCs are therefore not confined by external ISM pressure, which would allow them to be long lived while having $\ap \gg 1$ \citep[see][]{Heyer:2009aa,Roman-Duval:2010aa}.

Galactic scale simulations show that large scale ($\gtrsim 100 \kpc$) flows and drivers of turbulence are important components for GMC formation \citep[][]{Agertz:2009aa,Rey-Raposo:2015aa,Grisdale:2016aa}, which in the absence of a realistic galactic framework requires an explicit driving model \citep[e.g.][]{Saury:2014aa}. Galactic scale models allow for the exploration of how the location of a GMC within a galaxy (i.e. in arms or the bar etc.) affects its properties \citep[e.g.][henceforth F14]{Fujimoto:2014aa}, in addition to investigating the role of feedback \citep[e.g.][]{Hopkins:2012aa,Fujimoto:2016aa}. Such investigations have shown that feedback can act to limit the properties of a population of GMCs, in particular their mass and radius.

In recent years it has become possible to run simulations of entire disc galaxies reaching (sub-)parsec scale resolutions \citep[e.g.][]{Renaud:2013aa,Jin:2017aa}, encompassing self-consistently both the galactic and cloud scales, and thus simultaneously capturing the internal and environmental physics of GMCs. Following these lines, and expanding upon the work of \cite{Hopkins:2012aa} and \cite{Fujimoto:2016aa} \citep[see also][]{Baba:2017aa}, we detail here the role of feedback in determining the properties of a large number of GMCs in galactic context. We achieve this by identifying, using well tested 2D and 3D clump finders, and studying GMC populations in a MW-like simulation which is run both with and without feedback. Using these GMC we also study GMC scaling relations, with a particular attention to how the underlying distribution of the cloud virial parameter is linked to the inferred power-law exponents.

This paper is organised as follows: In  \S\ref{sect:meth}, we summaries the simulations and observational data used, as well as detail our GMC identification procedure. In \S\ref{sect:results}, we present the GMCs identified, compare them to observations, measure the Larson relations and explore different factors which may influence them. We discuss our findings in \S\ref{discussion} with respect to previous work. Finally, in \S\ref{con} we present our conclusions.
\begin{figure*}
	\begin{center}
		\includegraphics[width=1.0\textwidth]{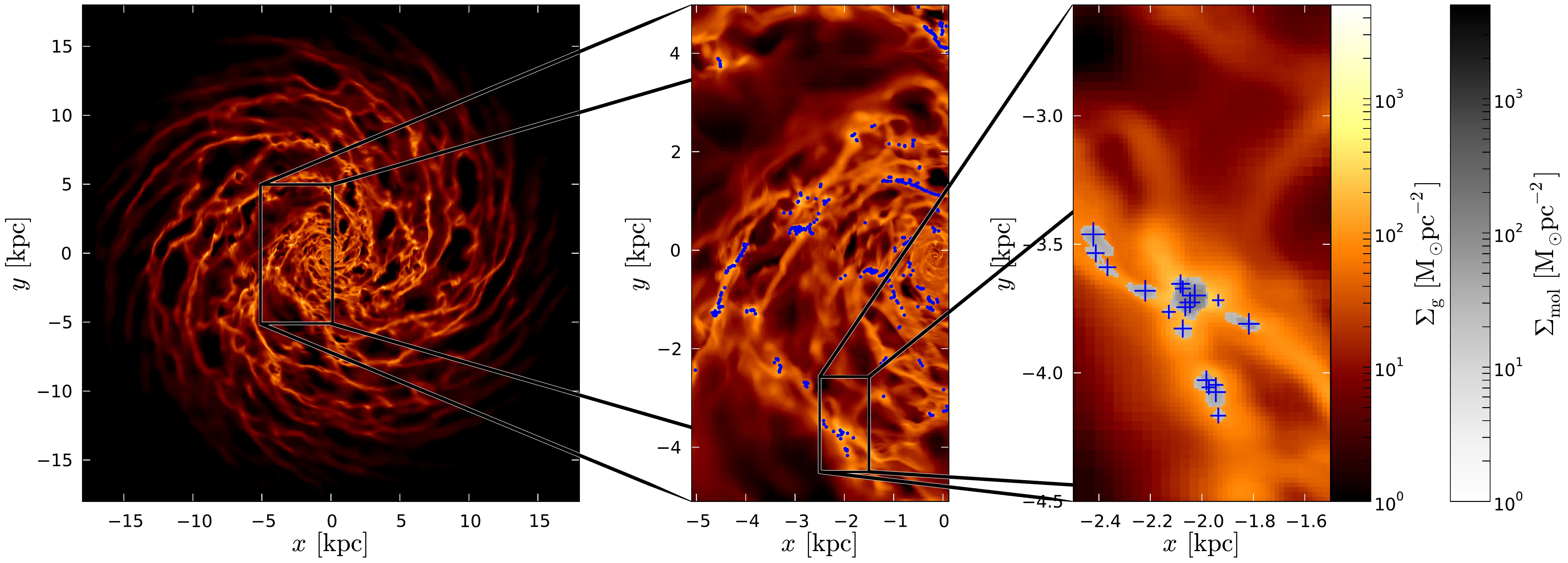}
		\caption{Left: Total gas surface density map of the MW simulation with feedback at $t=325\Myr$. Middle: Zoom-in of the total gas surface density and showing the GMCs identified by the 2D method (see \S\ref{method:2dclumps}) by the blue circles. Right: Zoom-in with an overly of the ``molecular'' gas used by the 2D clump finder and the position of GMCs (blue `+') in this region.  The size of the `+' indicates the relative size of each GMC. All three panels use the same surface density scale (shown on the right of the right panel). }
		\label{fig:galmap}
	\end{center}
\end{figure*}
 
\section{Method}
\label{sect:meth}

\subsection{Simulation suite}
\label{method:sims}
For this work we make use of the MW-like galactic disc simulations in \cite{Grisdale:2016aa} (henceforth G17). The simulations were run using  the hydro+$N$-body, Adaptive Mesh Refinement (AMR) code {\ramses} \citep{Teyssier:2002aa}. Feedback is implemented using the methods described in \bible and \cite{Agertz:2015ab}. Briefly, this feedback prescription includes the injection of  energy, momentum, mass and heavy elements over time via SNII and SNIa explosions, stellar winds and radiation pressure into the surrounding ISM. We follow the approach by \cite{Kim:2015aa} and inject momentum when a supernova cooling radius is unresolved, otherwise we inject thermal energy \citep[see][for details]{Agertz:2015aa}. \cite{Kim:2015aa} found that resolving the cooling radius with 3 grid cells was necessary to account for the correct momentum injection from a single supernova. We conservatively demand 6 grid cells, which is the case for $\sim50\%$ of SNe in our simulation.

The initial conditions (ICs) are the so called {\small AGORA} initial conditions \citep[][]{Kim:2014aa,Kim:2016aa}, and feature a stellar disc, stellar bulge, gaseous disc and dark matter halo. The particles distributions are set up following the approach by \cite{Hernquist:1993aa} and \cite{Springel:2000aa} \citep[see also][]{Springel:2005aa}, assuming an exponential surface density profile for the disc, a Hernquist bulge density profile \citep{Hernquist:1990aa}, and an NFW dark matter halo profile \citep{Navarro:1996ab}. Each simulation uses $10^6$ particles for both the NFW halo and stellar discs, with the same mass resolution in the bulge component as in the disc. The gaseous disc is initialised on the AMR grid assuming an exponential profile, as described in \cite{Agertz:2013aa}. The MW ICs were designed to have similar characteristics of a typical Sb-Sbc galaxy, like the MW.

Each galaxy is simulated in isolation (i.e. neglecting environmental effects such as galaxy interactions) and is embedded in a hot ($T=10^6\K$), tenuous ($n=10^{-5}\cc$) corona enriched to $Z=10^{-2}Z_\odot$, while the discs have the abundance  $Z = 1.5Z_\odot$. The galaxies are positioned at the centre of a simulation volume with a size of $L_{\rm box}=600\kpc$, and run with a maximum of 17 levels of adaptive mesh refinement, allowing for a finest grid cell size of $\Delta x\sim 4.6$pc.  The mass refinement threshold is $M_{\rm ref}\approx 9300 \Msol$.

Star formation in the simulations occurs according to the star formation law 
\begin{equation}
	\label{eq:schmidtH2}
	\dot{\rho}_{\star}=f_{\rm H_2}\frac{ \rho_{\rm g}}{t_{\rm SF}},
\end{equation}
where $f_{\rm H_2}$ is the local mass fraction of molecular hydrogen (H$_2$), $\rho_{\rm g}$ is the gas density in a cell, and $t_{\rm SF}$ is the star formation time scale of {\it molecular} gas. The fraction of molecular hydrogen in a cell is a function of the gas density and metallicity and is computed using the KMT09 model \citep[][]{Krumholz:2008aa,Krumholz:2009ab} implemented as described in \cite{Agertz:2015ab} (see their \S2.3, equations 2-6). $t_{\rm SF}$ is related to the {\it local} efficiency of star formation in a computational cell of density $\rho_{\rm g}$ by $t_{\rm SF}=t_{\rm ff, SF}/\epsilon_{\rm ff, SF}$, where $t_{\rm ff, SF}=\sqrt{3\pi/32G\rho_{\rm g}}$ is the local free-fall time of the star forming gas and $\epsilon_{\rm ff, SF}$ is the local star formation efficiency per free-fall time. 

We consider two galactic disc simulations, one with feedback and one without. For the simulation without feedback a local star formation efficiency per free-fall time of $\epsilon_{\rm ff, SF}=1\%$ is used. This low efficiency, motivated by the results of e.g. \cite{Krumholz:2007aa}, leads to a galaxy matching the empirical $\Sigma_{\rm SFR}-\Sigma_{\rm gas}$ relation \citep[][]{Kennicutt:1998aa,Bigiel:2008aa}, as shown by \cite{Agertz:2013aa}, and implicitly assumes regulated star formation, albeit without the explicit action of feedback. In contrast, in the feedback regulated simulation adopts  $\epsilon_{\rm ff, SF}=10\%$, i.e. allowing for feedback to regulate the star formation process back to the observed low efficiencies \citep[e.g.][]{Agertz:2016aa}. 

To limit the effect of artificial fragmentation \citep{Truelove:1997aa} we adopt a non-thermal Jeans pressure floor. To allow the local Jeans length to be resolved by $N_{\rm Jeans}$ cells of size $\Delta x$, the required non-thermal pressure is
\begin{equation}
P_{\rm Jeans}=\frac{1}{\gamma\pi}N_{\rm Jeans}^2G\rho_{gas}^2\Delta x^2,
\end{equation}
where $G$ is the gravitational constant and $\gamma=5/3$ is the adiabatic index. We adopt a Jeans number $N_{\rm Jeans}=10$. In the simulation with and without feedback, $\sim 0.1-0.5\%$ and $\sim 15-20\%$ of the gas mass has $P<P_{\rm Jeans}$ at any time, respectively. This equates to $\sim5-9\%$ and $\sim55-60\%$ of the total gas mass inside the GMCs identified with the 3D clump finder.

\subsection{GMC identification in 2D}
\label{method:2dclumps}
While observations of GMCs provide two-dimensional position data, as well as line-of-sight velocity information in so-called Position-Position-Velocity (PPV) space, simulators normally have access to three-dimensional position data (often referred to as Position-Position-Position, PPP, space).  Recent work comparing the properties of clouds detected in both PPV and PPP space have found only small variations for quantities that can be directly measured \citep{Khoperskov:2016aa,Pan:2015aa,Pan:2016aa}. However they did find that derived properties such as $\ap$ did show a more significant difference between the two methods.
The primary goal of this work is to understand the origins of observed GMC properties, such as mass, velocity dispersion and size, how they are correlated (Larson relations),  and the role that feedback plays. In order to compare simulations to observations, we identify GMCs both in projection, i.e. in 2D as described below, but separately carry out a full 3D clump finding analysis to understand how projected properties may differ from the actual ones. 

In order to identify GMCs in a manner consistent with observations we create 2D, face-on molecular gas surface density maps from the simulations. 
In G17, we found that the KMT09 model (see \S\ref{method:sims}) used for calculating the \HH content of our simulations was able to accurately reproduce the dense molecular environments of star formation but failed to reproduce the more extended \HH structures observed in galaxies, at least for our choice of parameters. Using the KMT09 model in this work results in only very small dense molecular clouds being detected, which would be more akin to observing only the densest inner regions of GMCs. As we are interested in the extended structures of GMCs we have instead opted to  assume that any gas with  $\rho\geq\rho_{\rm mol}$ is molecular, where $\rho_{\rm mol}$ a threshold density above which all gas is molecular. We adopt $\rho_{\rm mol}=100\cc$, which is well motivated given the gas metallicity adopted in this work \citep[][]{Krumholz:2009ab,Gnedin:2009aa}. We note that our choice of determining the molecular gas fraction for star formation and clump finding in different ways is not completely self-consistent. In future work we will explore this further, as well as how different star formation recipes impact GMC properties.

Before applying any clump finding algorithms we pre-process each simulation snapshot by removing all cells with $\rho<\rho_{\rm mol}$ and produce a surface density map of all the remaining ``molecular'' gas. We then employ the clump finding identification and analysis package \cupidtext, which is part of the Starlink Project \citep[see][for details]{Starlink2,Starlink} to identify GMCs in the molecular surface density maps. \cupid has several algorithms which can be used to identify GMCs. For simplicity and ease of comparison with observations we employ the 'clump find' method \citep{Williams:1994aa}, parametrised by a resolution in density contours ($\Delta C$), a density threshold for clumps ($C_{\rm min}$) and a minimum number of map pixels ($n_{\rm p,min}$) per clumps. \cupid first creates, at the maximum spatial resolution of the simulation ($\Delta x=4.6$ pc), a contour map of the gas surface density, with contour intervals of $\Delta C$. This map is in turn used to determine whether a given pixel is part of an over density.

In \cite{Rosolowsky:2007aa} and \cite{Ward:2016aa}, $\Delta C$ and to $C_{\rm min}$ are set to be multiples of the root mean square of the entire surface density map. We performed a parameter sweep to find the combination of $\Delta C$ and $C_{\rm min}$ which identifies GMCs in the feedback simulation with a similar range and distribution of masses ($10^{4}\leq M_{\rm GMC}<10^{7}\Msol$), radii ($R_{\rm GMC}\leq100\kpc$) and surface densities ($10\leq\Sigma_{\rm GMC}<10^{3}\Msol\,\pc^{-2}$) to those found in the MW in  \cite{Heyer:2009aa}. Finally , we adopt throughout this work $\Delta C=1.0\Msol\,\pc^{-2}$, $C_{\rm min}=10.0\Msol\,\pc^{-2}$ and $n_{\rm p,min}=9$. 

Fig.~\ref{fig:galmap} shows the surface density of gas in the simulated galaxy with feedback at $t=325$ Myr.
The left panel shows a large scale view of the disc, while the middle panel shows a zoomed-in portion, overlaid with circles indicating positions of GMCs identified with the 2D method. The right panel zooms in even further at an association of GMCs. Here we overlay the gas classified as ``molecular'' used by the clump finder, and indicate the relative sizes of GMCs. The strong shear and complex structure in the galactic innermost $\sim 0.5\kpc$ forbids a clear identification of over densities as clumps. We thus ignore any detection in this volume.

Once clumps have been identified, we are able to calculate their properties, including their mean surface density ($\Sigma$), mass ($M$), velocity dispersion ($\sigma$), virial parameter $\ap$, and the face-on surface area ($A$). We match the method used by observers (i.e. H09) to compute $R$, i.e. we assume spherical symmetry and calculate the cloud's radius from its measured $A$ as $R=\sqrt{A/\pi}$.
\begin{figure*}
	\begin{center}
		\includegraphics[width=0.85\textwidth]{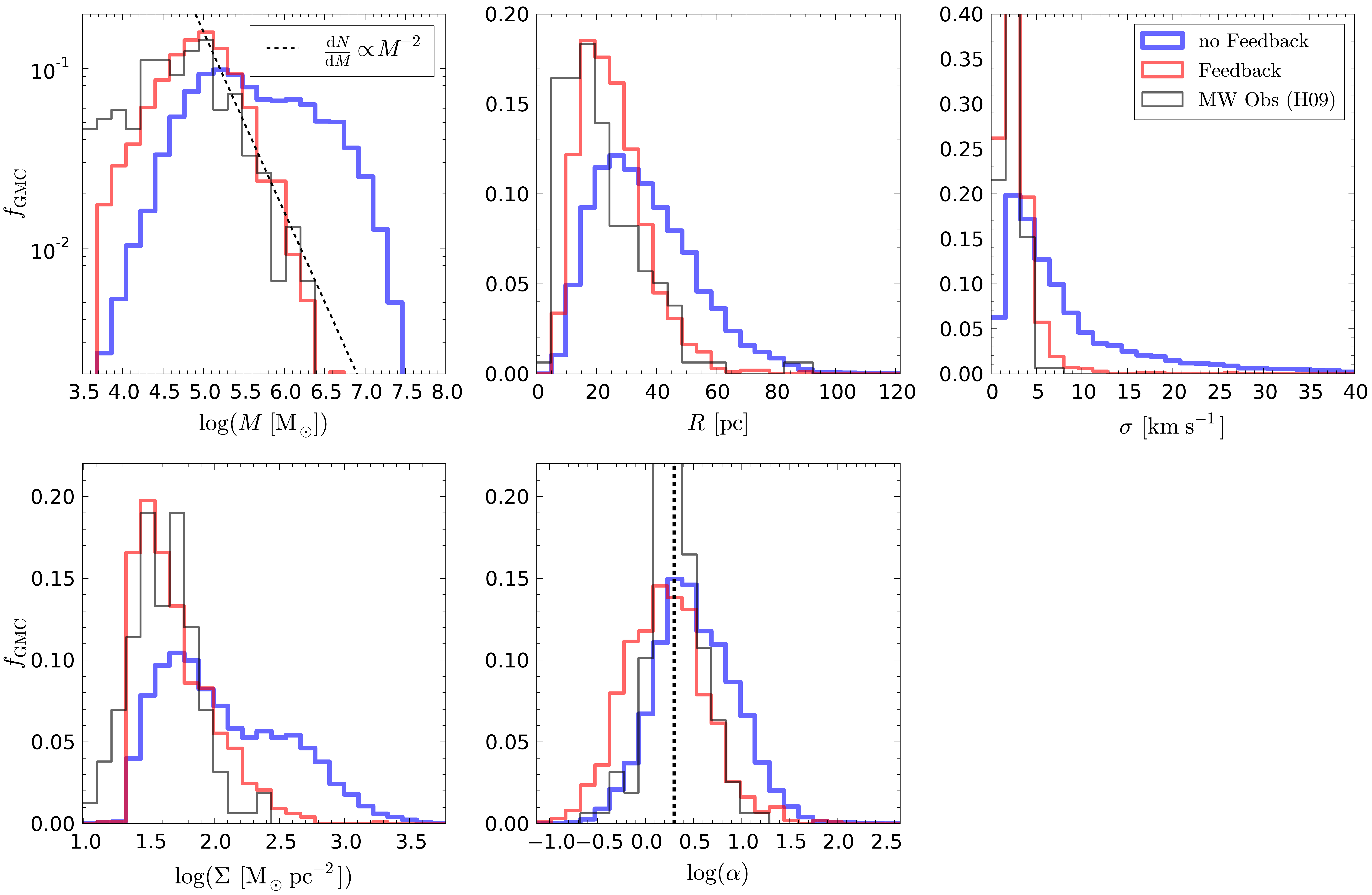}
		\caption{Normalised distributions of the properties (mass, radius, velocity dispersion, surface density, virial parameter) of GMCs detected with the 2D method. The red, blue and black lines show data from the simulations with feedback, without feedback and data from H09 respectively. The dashed vertical line in the middle-bottom panel shows $\ap=2$, i.e the border between bound and unbound. 		}
		\label{fig:2D_hist}
	\end{center}
\end{figure*}
\begin{figure*}
	\begin{center}
		\includegraphics[width=0.95\textwidth]{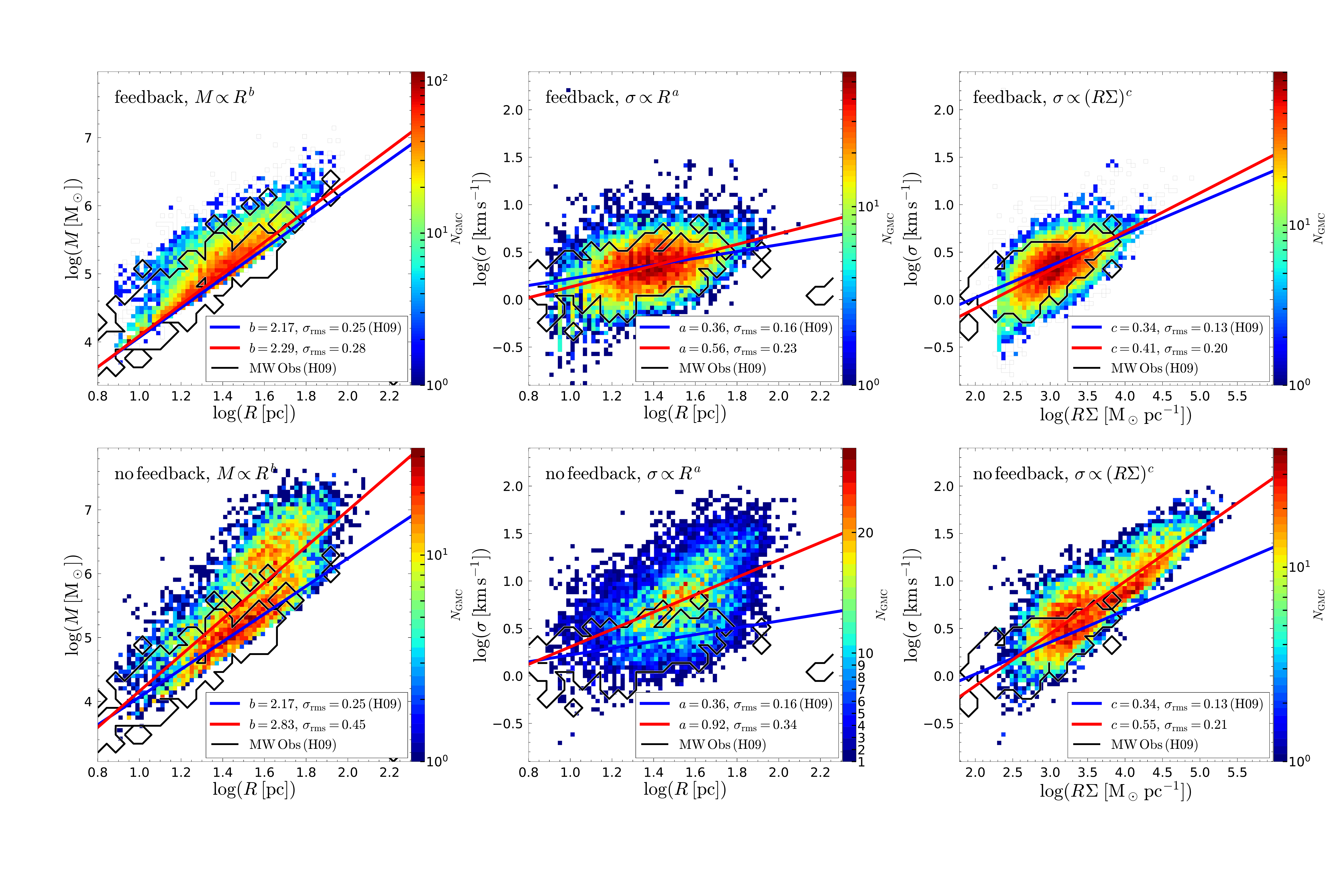}
		\caption{2D histograms comparing properties of the 2D GMCs. From left to right we show the mass of GMCs ($M$) as a function of their radius ($R$), their velocity dispersion ($\sigma$) as function of $R$ and $\sigma$ as a function of $R\Sigma$ where $\Sigma$ is the surface density of the GMC. The top row shows the data for GMCs found in the feedback simulation, while the bottom row shows GMCs in the simulation without feedback. We fit a Larson-like relation for each panel (red line) and show the relationship we measure from H09's data (blue line). The measured power law and the root-mean-square scatter ($\sigma_{\rm rms}$, in dex) of these two fits are given in each panel.  The black contours show the equivalent values for observed MW GMCs, as measured by H09.  
		}
		\label{fig:2D_larson}
	\end{center}
\end{figure*}

\subsection{GMC identification in 3D}
\label{method:3dclumps}

To connect the observed projected characteristic of clouds to their three-dimensional properties, we make use of the on-the-fly 3D clump finding module built into \ramses: PHEW \citep[Parallel HiErarchical Watershed, see][for details]{Bleuler:2014aa}. PHEW first identifies AMR cells above a given density threshold ($\rthres$). Nearby dense cells are considered part of the same clump when the ratio of their densities is less than a relevance parameter $r$. Finally, adjacent clumps separated by a density valley larger than a saddle density parameter $\sthres$ are merged, while those with deeper density differences remain different objects. We adopt $\rthres=\rho_{\rm mol}=100\cc$, $r=1.2$ and $\sthres=104\cc$, to match a visual inspection of the clouds in our simulations. We checked that changing these parameters have little impact on our conclusions. As in the 2D approach we neglect all GMCs with $\rg\leq0.5\kpc$. Cloud sizes are computed using the total volume $V$ of their cells assuming spherical symmetry, i.e. $R = \sqrt[3]{(3V)/(4\pi)}$.

\subsection{Observational data}
\label{method:obs}
To ensure that the objects we identify as GMCs are similar to those found in observations, part of our analysis involves comparing  properties of  GMCs identified in simulations to observations. We do this by comparing clouds identified in projection (2D) to properties of GMCs measured by H09. In their work they re-analysed 162 of the GMCs in the \Sol catalogue and used `The Boston University-FCRAO Galactic Ring Survey' (GRS) to supplement the original observations with higher quality data. The GRS observations provided higher angular sampling and spectral resolution. It used the mostly optically thin $^{13}$CO tracer which reduced the effect of velocity crowding and provided details of internal cloud structure. For full details of the observations we refer the reader to H09 and the references within. 

We make use of the data for each GMC present in table 1 of H09. For several GMC properties they provide two sets of values: A1 and A2 values. The A1 data set uses the same GMC areas as found by \Sol while A2 data uses an area defined  as the area within the half-power isophote of the peak column density value within the cloud. As the second data set is analogous to the cores of GMCs we make use of the A1 data set. As with our simulated GMCs we treat the GMCs from H09 as if they where spherical when determining their radius.

\begin{figure*}
	\begin{center}
		\includegraphics[width=0.9\textwidth]{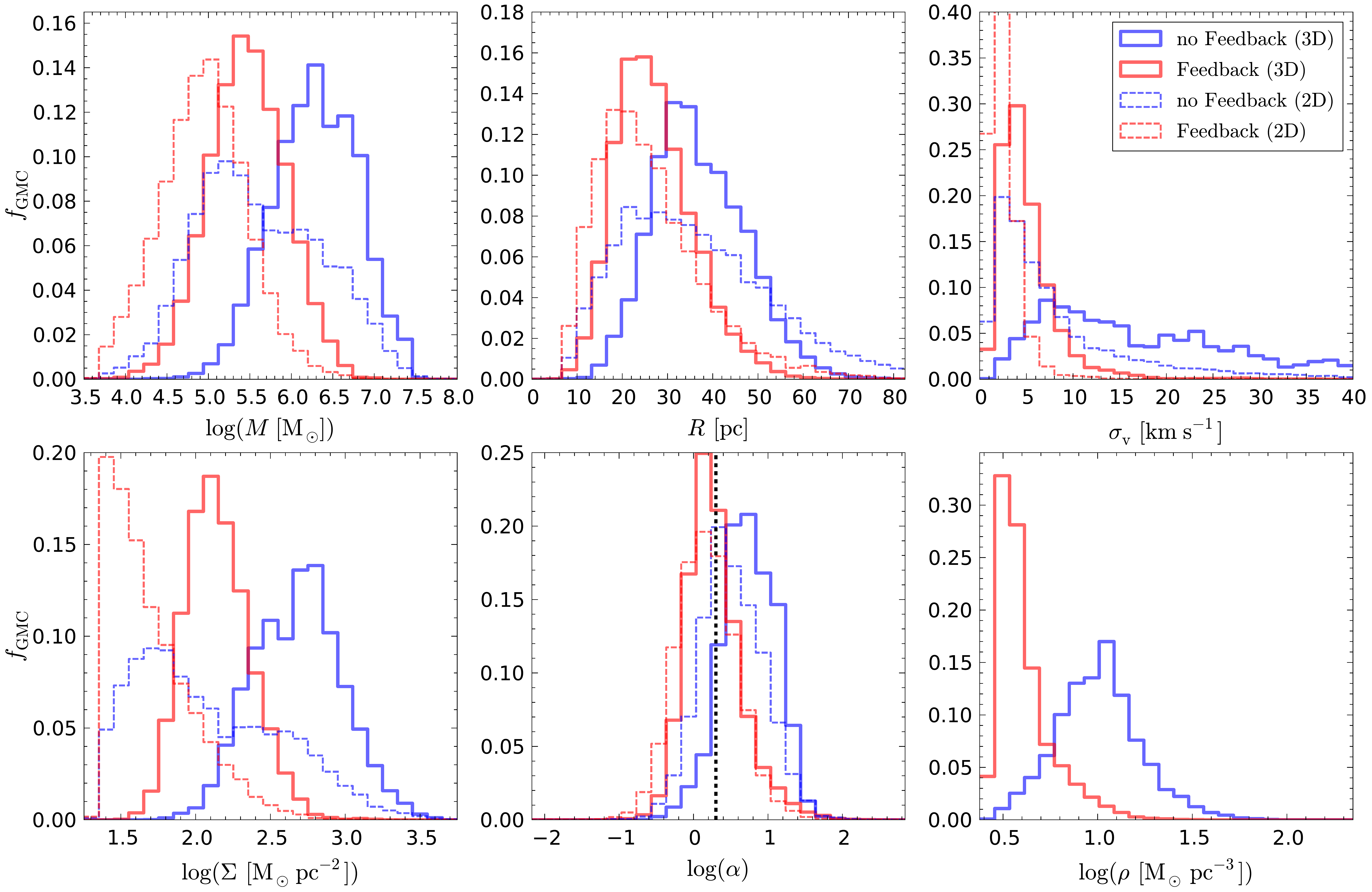}
		\caption{Same as Fig.~\ref{fig:2D_hist}, but comparing the GMCs detected in 2D and in 3D. The bottom-right panel shows the distribution of volume density of clouds. 
		}
		\label{fig:3D_hist}
	\end{center}
\end{figure*}

\begin{figure*}
	\begin{center}
		\includegraphics[width=0.9\textwidth]{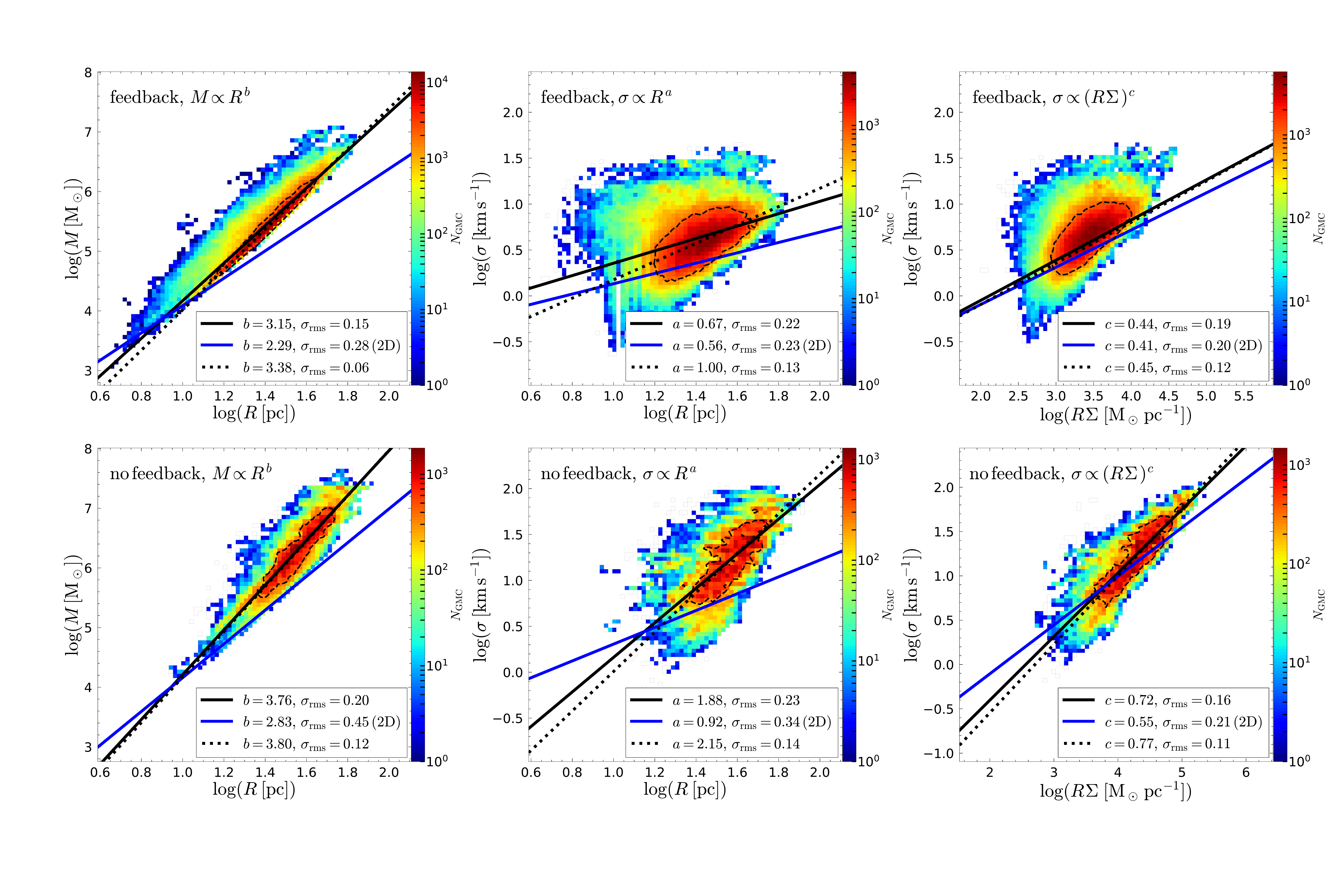}
		\caption{Same as Fig.~\ref{fig:2D_larson}, but for our 3D GMCs. We fit a Larson-like relation for each panel (solid black line) and compare with the 2D equivalent (blue line). The dashed black line shows an additional fit to each distribution, but limited to the most populated region, as indicated by the dashed black contour (see text). As before we give root-mean-square scatter ($\sigma_{\rm rms}$, in dex) for each fit
		}
		\label{fig:3D_larson}
	\end{center}
\end{figure*}
\section{Results}
\label{sect:results}

In our previous work, G17, we modelled an isolated MW-like galaxy, both with and without feedback. We demonstrated  that the model with feedback 
was able to produce  a turbulent ISM that matched observations on scales from $\sim40\pc$ up to several $\kpc$ (see Fig.~ 5 and 9 of G17). In this work we continue the analysis of these simulations but focus on how feedback affects the structure and properties of the ``molecular'' component of the ISM on scales up to a few $100\pc$.

\subsection{GMCs in 2D}
\label{results:2d}
In the following section we present results of the 2D clump finding method described in \S\ref{method:2dclumps} and compare the simulated GMCs to observations. 

\subsubsection{General Properties}
\label{results:2d:gen}
In Fig.~\ref{fig:2D_hist} we show the normalised distribution of GMC masses, radii, velocity dispersions, surface densities and viral parameters found in our simulations with and without feedback, as well as observations from H09. The analysis is carried out once the simulation has reached a steady phase of galaxy evolution, i.e. with a spiral galaxy morphology and a roughly constant star formation rate (see G17 for details). The normalised distributions of clouds at any time during this phase are very similar, with the same mean and spread in values. We therefore stack data from 12 different snapshots, taken over $300 \Myr$, for better statistics.

We find a clear distinction between the two simulations, with feedback preventing GMCs from having masses, radii and velocity dispersions greater than a few $10^{6}\Msol$, $\sim70\pc$ and $10$-$15\kmsec$ respectively. With feedback, the distributions are in very good agreement with the observation by H09, although we note that the low mass/small radii end of distributions are uncertain due to both observational and numerical limitations. In particular, the H09 data feature a population of low mass clouds ($10^{3.5}\Msol\lesssim M\lesssim10^{4}\Msol$) which we do not resolve in the simulations.

For $M>10^{5}\Msol$, the mass function of MW clouds and the feedback model are in excellent agreement, where both follow a power law distribution with an index of $\sim -2$. This is in line with power law indices derived from other MW GMC surveys, as well as extra-galactic surveys \citep[see][and references within]{Dobbs:2014aa}. 

GMCs in the simulation without feedback either form with (or grow) up to masses up to $\sim10^{7.5}\Msol$, radii extending to $\sim 100\pc$, and velocity dispersions as large as $\sim 100\kms$ (not visible in the figure). This leads to a larger fraction of unbound clouds ($\ap>2$), and the distributions from this run are in poor agreement with observations.

The properties of a GMC are dependent on the resolution of the simulation, in particular the distributions of mass and radius (see Appendix~\ref{App:resolution}). As a result in order to recover the observed distributions simulations need to be able to resolve the radius of a GMC with several cells, as we do in our ($\Delta x\sim4.6\pc$) simulations. However, note that despite the excellent agreement between our simulations and observations we do not know if our results have converged. Higher numerical resolution is required to settle this question, which we leave for future work. 
\subsubsection{GMC scaling relations (2D)}
\label{results:2d:larson}

Fig.~\ref{fig:2D_larson} shows the distributions of our (stacked) clouds in the planes of the Larson relations. In each panel, we show our simulated GMCs (heat map) and the position of all the observed clouds from H09 (black contours), and linear least squares fit to our simulated data (red line) to the H09 GMCs (blue line). We note that due to the resolution limit of our simulations, and the requirement that all GMCs have a minimum of 9 cells, we are unable to study the scaling relations for sizes $\lesssim8\pc$. 

In all three parameter spaces, we retrieve the Larson correlations, with $a\sim0.56$, $b\sim 2.29$ and $c\sim 0.41$ (see Eq.~\ref{eq:larson1}--\ref{eq:larson3}) in the simulation with feedback. This is in excellent agreement with the H09 data, as well as other recent work discussed in \S\ref{sect:intro} \citep[e.g. $b=2.36\pm 0.04$,][]{Roman-Duval:2010aa}, with a highly significant overlap between observations and our most densely populated region. The simulations feature larger scatter around the mean relations compared to observations, which is likely an effect of stacking which multiplies the number of detection of short-lived clouds on the verge of dispersing. 

Without feedback, the scaling relations are always steeper, in poor agreement with the H09 data, particularly in the $\sigma$--$R$ and $\sigma$--$R\Sigma$ planes. This adds further evidence that feedback play an important role in the formation and shaping of GMCs.

\subsection{GMCs in 3D}
\label{results:3d}

\subsubsection{General properties}
\label{results:3d:gen}

In Fig.~\ref{fig:3D_hist}, we show the normalised distributions of GMCs, this time identified with the 3D clump finder (see \S\ref{method:3dclumps}). Here again, we stack the distributions from several snapshots to eliminate small-number statistics.

The differences between GMCs in the simulation with and without feedback found in projection is also present when analysed in 3D. As before we find that feedback acts to limit the masses, sizes, velocity dispersions, surface densities and virial parameters. The same is true for their physical densities, as shown in the bottom right panel; in the feedback simulations GMCs are less dense, with $\sim75\%$ of GMCs being found with $10^{0.5} \Msol\pc^{-3}<\rho<10^{0.8}\Msol\pc^{-3}$. Without feedback we find that $\sim80\%$ of GMCs are found with $10^{0.8} \Msol\pc^{-3}<\rho<10^{1.3}\Msol\pc^{-3}$. 

We find an almost uniform shift in the 3D property distributions compared to their 2D counterparts, with the 3D properties being shifted to larger values in all cases (except $\ap$). Despite these differences, the maximum cloud mass in the two detection methods is the same ($\sim 10^{7}\Msol$). As we are using the same definition of molecular gas in both methods ($\rho_{\rm mol}=100\cc$), the cause of this difference is likely due to projection effects. Isolated regions of molecular gas that wouldn't be identified as GMCs by the 3D clump finder (e.g. too few connected cells) can in projection appear as coherent, but low mass, GMCs \citep[see also][]{Duarte-Cabral:2016aa}. 

As mentioned above, it is not straightforward to directly compare observed GMC properties to those computed from the 3D clump finding method. However, given our excellent match between the simulated GMCs detected in projection and observed GMCs, we speculate that observations in principle also may suffer from the similar projection effects, with similar biases towards higher frequencies of low mass GMCs.

The resolution of the simulation also effects properties of the GMCs identified using the 3D clump finder, see Appendix~\ref{App:resolution}. As with the 2D clump finding, it is the mass and size of the cloud that are most strongly affected. With sufficiently high resolution simulations (i.e. $\Delta x\lesssim1\pc$) it is probable that the properties of the GMCs will become resolution independent. However at such resolutions clump finders will need to be able to distinguish between small GMCs and substructures within the same cloud. 
\begin{figure*}
	\begin{center}
		\includegraphics[width=0.95\textwidth]{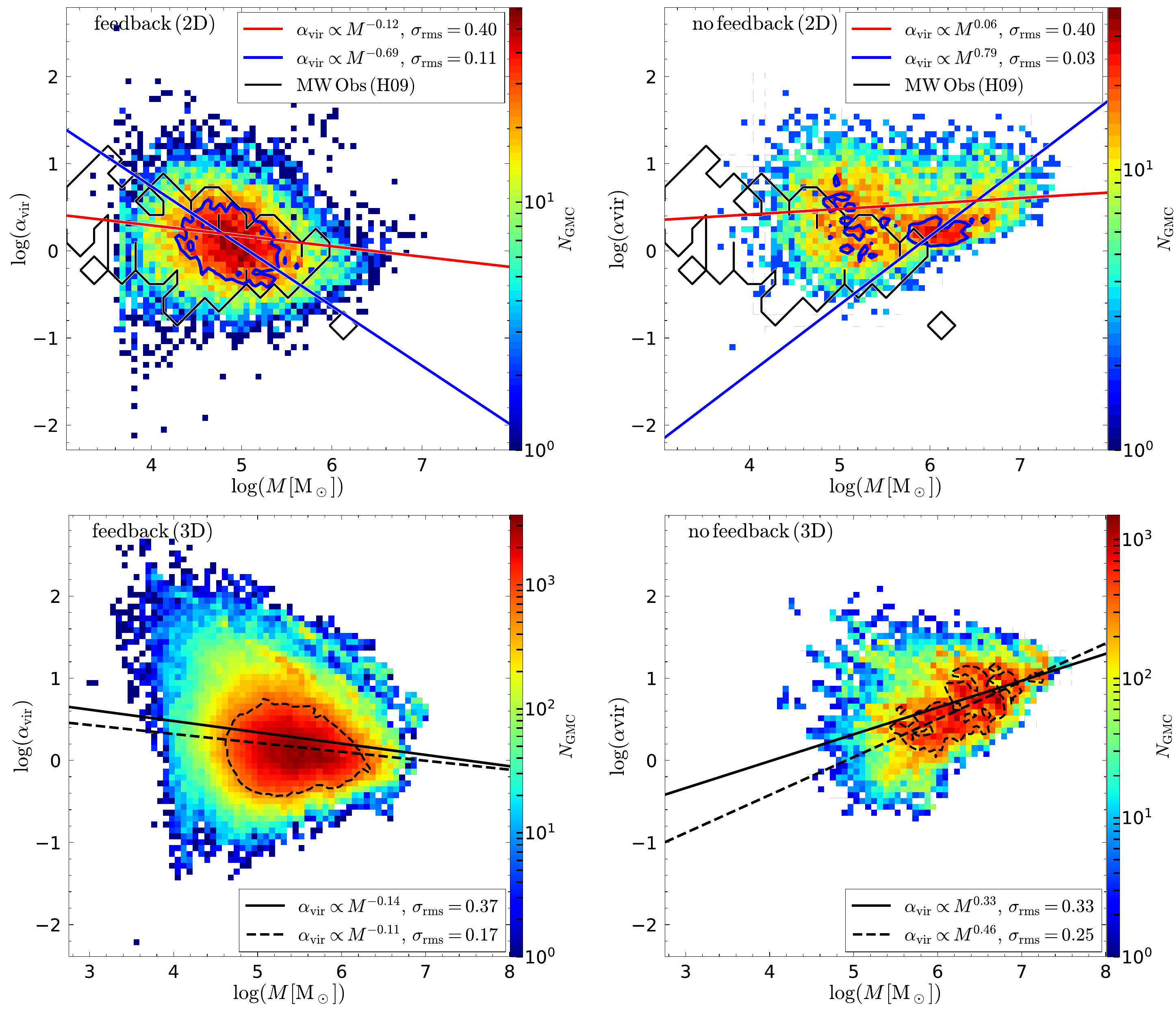}
		\caption{Distributions of the GMCs' virial parameters ($\ap$) as a function of their mass (M), with (left) feedback and without (right) feedback. The top row shows GMCs detected using the 2D clump finder (heat map) with the data from H09 superimposed (black contours). The bottom row shows GMCs detected using the 3D clump finder. In all panels we fit a power law to all the data (solid black line) and to densest regions only (solid blue line and dashed black lines), i.e. the area described by the blue contour (top row) and black-dashed contour (bottom row). The root-mean-square scatter ($\sigma_{\rm rms}$, in dex) for these fits is also given. 
		}
		\label{fig:3D_apvmass}
	\end{center}
\end{figure*}
\subsubsection{GMC scaling relations (3D)}
\label{results:3d:larson}

In Fig.~\ref{fig:3D_larson}, we show the same scaling relations as in Fig.~\ref{fig:2D_larson}, but now using cloud properties obtained from the 3D clump finder. 

The shift from 2D to 3D only produces a marginal change in of $a$ and $c$ for GMCs in the simulation with feedback: $a$ goes from $0.56$ to $0.67$ and $c$ changes from $0.41$ to $\sim0.44$. However, $b$ steepens significantly from $2.29$ to $3.15$. 
The steepening of the $M$-$R$ relation is expected when volume density thresholds are used to define clouds \cite[see][]{Ballesteros-Paredes:2012aa} and thus is also expected for $a$ and $c$ (see \S\ref{results:boundness}).

There is significant scatter in the simulated GMC scaling relations. By analysing only the 20\% most populated bins in the heat map of Fig.~\ref{fig:3D_larson} (indicated by the black-dashed contour) we calculate a second fit for each relation (dashed black line). We find that in the case of the $M$--$R$ relation and the $\sigma$--$R\Sigma$ relation, there is little difference between this fit and that from the full cloud population. However, the $\sigma$--$R$ relation steepens significantly from $a=0.67$ to $\sim 1$ in the case with feedback.

\subsection{The virial parameter and its relation to scaling relations}
\label{results:3d:alpha}
\begin{figure*}
	\begin{center}
		\includegraphics[width=1.\textwidth]{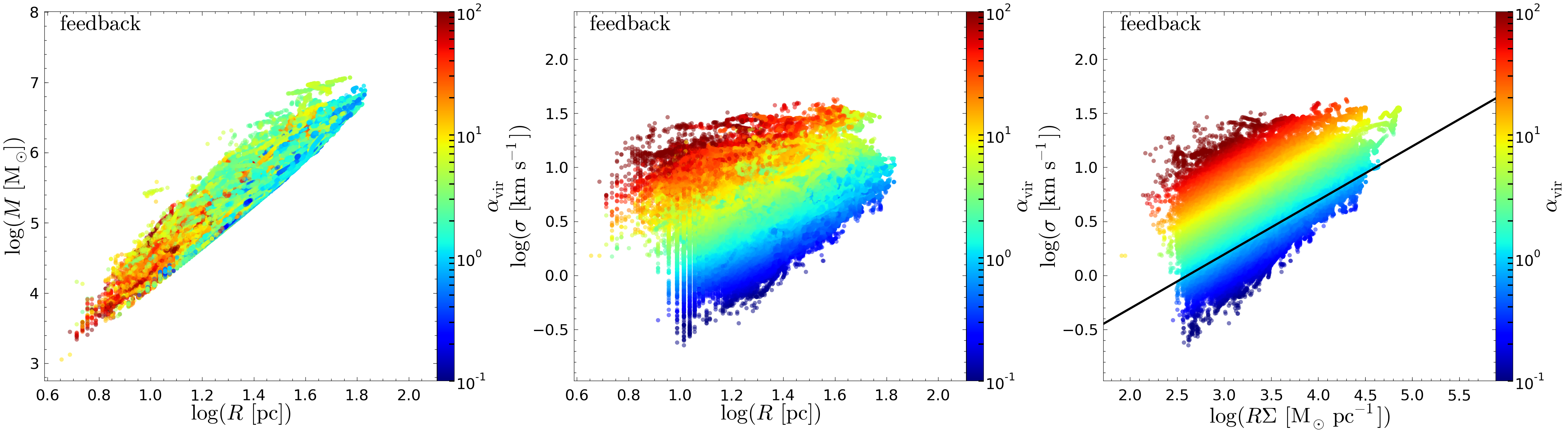}
		\caption{From left to right we show the mass of GMCs ($M$) as a function of their Radii ($R$), their velocity dispersion ($\sigma$) as function of $R$ and $\sigma$ as a function of $R\Sigma$ where $\Sigma$ is the surface density of the GMC in the simulation with feedback detected for using the 3D method (see \S\ref{method:3dclumps}). Each GMC is plotted individually with the colour of the point indicating the clouds virial parameter.  We also show the predicted relation between $\sigma$ and $R\Sigma$ for a population of clouds all with the same virial parameter $\ap=1$(,see Eq.~\ref{eq:lf5})
		}
		\label{fig:3D_larson_alpha}
	\end{center}
\end{figure*}
The virial parameter is thought to be a crucial factor in controlling the efficiency of star formation in GMCs \citep[e.g.][]{Padoan:2012aa,Rey-Raposo:2017aa}. Understanding and reproducing the observed distribution of $\ap$ (see Fig.~\ref{fig:2D_hist}), as well as how $\ap$ correlates with other GMC properties, is hence an important component in galaxy evolution modelling.

In Fig.~\ref{fig:3D_apvmass}, we show the relationship between mass and $\ap$ for GMCs identified using both the 2D (top) and 3D (bottom) methods. Assuming $\ap\propto M^{d}$ we find $d<0$ in the simulation with feedback and $d>0$ without. This suggests an interesting dichotomy: with feedback, massive clouds are more gravitationally bound, in contrast to models neglecting feedback where the majority of clouds of mass $\gtrsim 10^6\Msun$ have $\ap>1$, reaching $\ap\sim 10$ at $10^7\Msun$. This is consistent with numerical work by \cite{Fujimoto:2016aa} \citep[see also][]{Dobbs:2011aa} who adopted inefficient thermal feedback (see \S\ref{discussion}) and found $d>0$. GMCs in our model without feedback are long lived and feature higher velocity dispersions than those in the feedback driven models, possibly due to cloud-cloud interactions \cite[][]{Tasker:2009aa,Dobbs:2011aa} and hence have larger $\ap$. In future work, we will quantify how properties of individual GMCs evolve in time to better understand this discrepancy (Grisdale et al. in prep).

In projection, the $\ap$--$M$ relation from the simulation with feedback is in good agreement with MW observations (H09, black contour in the top row of Fig.~\ref{fig:3D_apvmass}). Fitting the entire simulated GMC population (black line) in the top left panel yields $d=-0.12$, is close to that of the MW data, where $-0.21$. By considering only clouds in the most populated region, indicated by the blue contours, of the $\ap$--$M$ plane\footnote{using the same definition as in \S\ref{results:3d:larson}} we measure a steeper relation, $d=-0.69$. While this steeper relation does not match the H09 data well it is a closer match to what was found for the galaxy-wide sample by MD17 ($d=-0.53$).

\subsubsection{Boundness and the scatter in GMC scaling relations}
\label{results:boundness}
To further understand the role of boundness for GMC properties we considering a cloud in virial equilibrium with its environment. For such clouds,
\begin{equation}
	2T + \Omega = 4\pi R^{3}\pv,
	\label{eq:lf1}	
\end{equation}
where $T={3M\sigma^{2}}/{2}$ is the kinetic energy of the cloud, $\Omega = {-3GM^{2}}/{5R}$ is the potential energy due to gravity\footnote{We adopt the mass profile of a homogeneous sphere for simplicity.} and $\pv$ is the external pressure required to keep the system in virial equilibrium. We thus have 
\begin{equation}
	3M\sigma^{2}+ \frac{-3GM^{2}}{5R} = 4\pi R^{3}\pv.
	\label{eq:lf2}
\end{equation}
Rearranging to get $\sigma$ on the left hand side and defining $\Sigma = \frac{M}{\pi R^{2}}$ we get
\begin{equation}
	\sigma [{\rm km\,s^{-1}}] = 0.05 \left(\frac{R\Sigma }{\Msol\,\pc^{-1}}\right)^{0.5} \left(1+\frac{6.67\pv}{\pi G\Sigma^{2}}\right)^{0.5}.
	\label{eq:lf3}
\end{equation}
Eq.~\ref{eq:lf2} can also be rearranged to give
\begin{equation}
	\frac{5\sigma^{2}R}{GM} =  1+\frac{6.67\pv}{\pi G\Sigma^{2}}.
	\label{eq:lf4}
\end{equation}
where the left hand side of Eq.~\ref{eq:lf4} is the virial parameter $\ap$. This allows Eq.~\ref{eq:lf3} to simplify to 
\begin{equation}
	\sigma [{\rm km\,s^{-1}}] = 0.05 \sqrt{\ap}\left(\frac{R\Sigma }{\Msol\,\pc^{-1}}\right)^{0.5}.
	\label{eq:lf5}
\end{equation}

In Fig.~\ref{fig:3D_larson_alpha} we show how $\ap$ affects the scatter in our simulated GMC scaling relations. The effect on the $\sigma$-$\Sigma R$ relation in the right panel is trivial, as seen from Eq.~\ref{eq:lf5} above, where $\ap$ sets the normalisation of the relation. We find a similar trend for the $\sigma$--$R$ relation, which indicates that $\ap$, with a $\sim2$ dex spread in values (Fig.~\ref{fig:2D_hist}), is the main driver behind the scatter. The spread in $\Sigma$ is only $\sim$1 dex and cannot alone account spread in $\sigma(R)$ \citep[][]{Ballesteros-Paredes:2011aa}. In the $M$--$R$ plane the separation between clouds with different $\ap$ is less clear, but clouds with $\ap<1$ tend to form a shallower relation. 

The above analysis show that the underlying $\ap$ distribution of an observed or simulated GMC population can affect the values of the power-law exponents in the Larson relations. Understanding GMC selection biases, e.g. due to incompleteness effects, sensitivity limits in observations, or clump finder settings, is therefore important. This point can be illustrated by rearranging Eq.~\ref{eq:lf5} to express $\sigma$ as a function of $M,\,\ap$ and $R$, and by replacing $\ap$ with $M^{d}$, hence relating the power law exponents $a$, $b$ and $d$ as 
\begin{equation}
	a = 0.5[b(1+d)-1].
	\label{eq:lf6}
\end{equation}
If we, for example, take a population of clouds where $d=0$ (i.e. all clouds have the same $\ap$) and $b=2$, then  Eq.~\ref{eq:lf6} implies $a=0.5$, and we have recovered the `classical' Larson scalings \citep[as found by][]{Larson:1981aa,Solomon:1987aa}. Similarly, for steeper $\ap$-$M$ relations, where $d>0$ as predicted by simulations without feedback, $a$ increases to values not compatible with Milky Way observations. 

\begin{figure*}
	\begin{center}
		\includegraphics[width=1\textwidth]{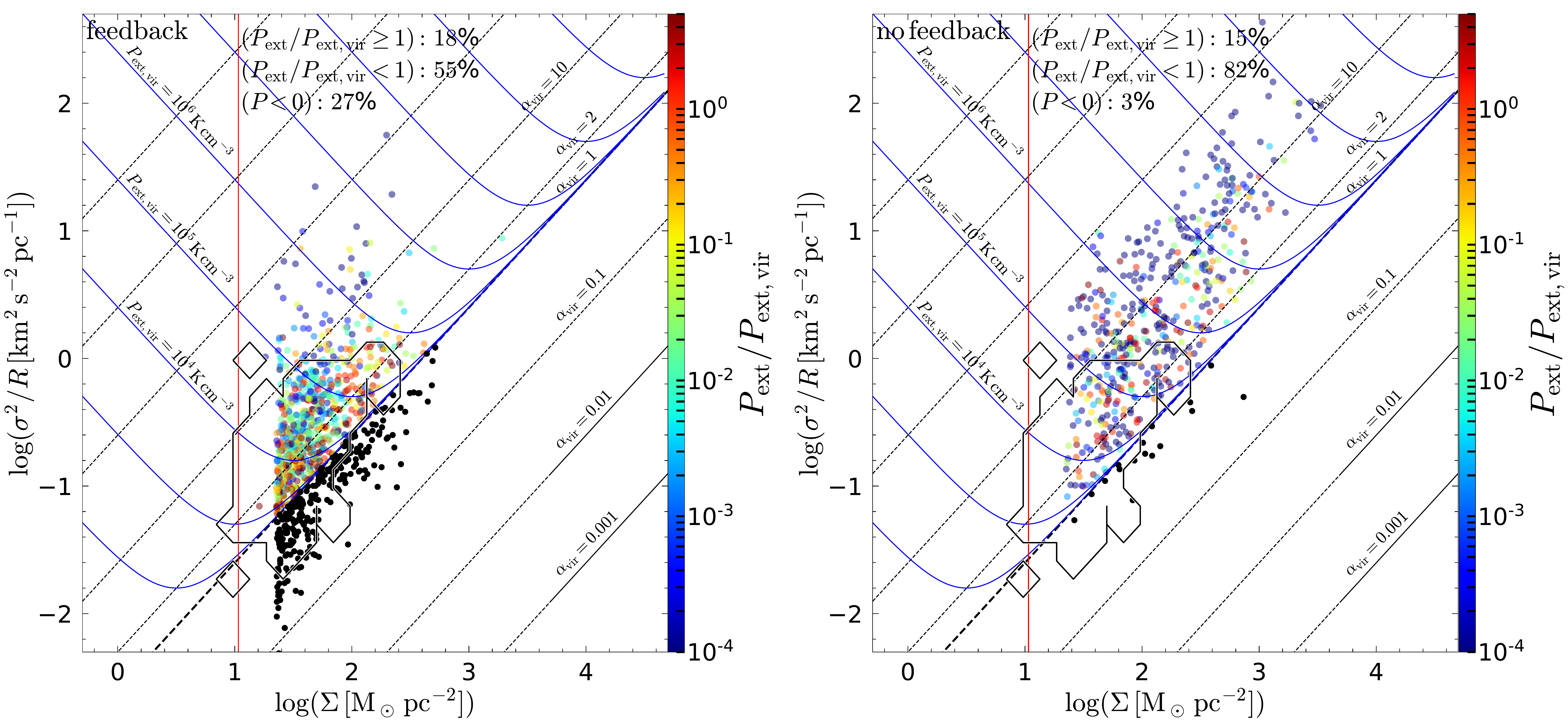}
		\caption{
		Position in the $\sigma^{2}/R$--$\Sigma$ plane for GMCs found using the 2D clump finder in the simulation with (left) and without (right) feedback. The colour each circle shows the ratio of $P_{\rm ext}$ to $\pv$. Black points indicate collapsing clouds ($\pv<0$). Overlaid are lines of constant $\ap$ (dashed-black) and $\pv$ (solid-blue). The vertical red-line indicates the detection limit of our clump finding technique. In both figures we only show data from a single snapshot for ease of reading. Data from H09 is overlaid as the black contour.
		}  
		\label{fig:leroy}
	\end{center}
\end{figure*}

\subsubsection{Bound by pressure or gravity?}
In the previous sections, we have evaluated the virial parameter of the clouds by only considering their internal properties, and neglecting environmental effects. However, considering external pressure on the clouds corresponds to shifting the levels of reference of $\ap$, such that a compressed cloud could \emph{effectively} be in virial equilibrium even with $\ap \gg 1$.

To understand the importance of external pressure, we assume it can be approximated by the galactic mid-plane pressure. Following \cite{Elmegreen:1989aa}, we thus thus measure
\begin{equation}
P_{\rm ext}=\frac{\pi}{2}G\Sigma_{\rm g}\left(\Sigma_{\rm g}+\Sigma_\star\frac{\sigma_{\rm g}}{\sigma_\star} \right),
\end{equation}
where $\Sigma_g$ and $\Sigma_\star$ are the surface densities of gas and stars, and $\sigma_{\rm g}$\footnote{both  thermal and turbulent motions are accounted for when computing the effective dispersion $\sigma_{\rm g}$} and $\sigma_\star$ are the corresponding velocities dispersions. These quantities are measured on $100 \pc$ scales at the location of each GMCs discussed in \S\ref{results:2d}. 

Fig.~\ref{fig:leroy} shows the distribution of clouds in the $\Sigma$--$\sigma^2/R$ plane (i.e. the two opposing effects potentially balancing each other in the case of equilibrium) at $t=325$ Myr, indicating both their virial parameters (dotted lines) and the external pressure they experience (colour of the points). This pressure $P_{\rm ext}$ is normalised to the external pressure $\pv$ (computed from Equation~\ref{eq:lf3}) the cloud would need to be in equilibrium, i.e. the external effect needed to restore equilibrium for a cloud that is otherwise unbalanced if considering internal processes only. Out-of-equilibrium, self-collapsing clouds yield $\pv<0$ and are marked as black dots. 

We define 3 regimes for GMCs based on the ratio of $P_{\rm ext}$ and $\pv$: $P_{\rm ext}/\pv\geq1$ (compressed by external pressure), $0\leq P_{\rm ext}/\pv<1$ (not confined by external pressure) and $\pv<0$ (undergoing gravitational collapse). The percentage of clouds in each of these regimes at $t=325$ Myr are shown in Fig.~\ref{fig:leroy}.  At all times in the simulation with feedback, the fraction of compressed, dissolving and collapsing clouds are  $\sim 15-25\%$,  $\sim 50-60\%$ and $\sim 25-30\%$, respectively. Without feedback, the fraction of clouds undergoing collapse at any time is only a few percent, with the majority of clouds not experiencing any significant external pressure ($0\leq P_{\rm ext}/\pv<1$).

As we consider clouds over the entire galaxy, the $\Sigma$--$\sigma^2/R$ plane in Fig.~~\ref{fig:leroy} encompasses a variety of objects, at various stages of their evolution, including non-star forming clouds, collapsing clouds, clouds being dispersed by dynamics (e.g. shear) and clouds being dispersed by internal feedback. As a result, it is complicated to identify clear patterns or evolutionary tracks. Not surprisingly, the diversity of external pressures translates into a range a physical properties, even for a given $\ap$. Monitoring the evolution of individual clouds in such a diagram will be presented in a forthcoming paper (Grisdale et al. in prep). We can however highlight trends.

The rather regular structure of our modelled galaxies (i.e. the absence of grand-design spirals, or bars) does not allow us to probe a large range of environmental conditions, as in e.g. \citet{Leroy:2015aa}. The lack of cosmological context (gas accretion, evolution of the external part of the disc(s), flaring etc.) is likely responsible for the absence of the family of very diffuse clouds in the outer galaxy, identified by \citet[see also \citealt{Leroy:2015aa}]{Heyer:2001aa}. 

Our simulation thus mainly comprises self-gravitating clouds $\ap \lesssim 2$, with the important additional degree of freedom offered by external pressure. Therefore, a significant number of clouds with high $\ap$, but confined by external pressure, can be detected in the simulation. This duality between self-bound and confined by external pressure could in principle lead to two regimes of cloud evolution, but their overlap make them virtually indistinguishable in the observable quantities we analysed. It is however possible that they feature different cloud lifetimes (Grisdale et al., in preparation).

Although lacking an estimate of timescales for the different phases discussed above, Fig.~\ref{fig:leroy} illustrates that the dispersions found in the scaling relations of GMCs (recall the previous section) in part is due to variation of external pressure, in addition to $\ap$, across the galactic disc. We can thus extrapolate that significantly stronger dispersions would be found in more extreme environments, such as galaxy mergers or gas rich high redshift galaxies.

\section{Discussion}
\label{discussion}
Over the past decade there have been a number of numerical studies aimed at exploring the origins of GMC properties. We now turn our discussion to how those studies relate to the work presented here.

Recent work by \cite{Tasker:2009aa} and \cite{Fujimoto:2014aa} have shown that GMC populations forming in global disc simulations, where ISM properties are set only by gravity and hydrodynamics (e.g. shear) and without feedback do not match observations. GMCs tend to be too massive, too large and have too high velocity dispersions. In addition, GMC scaling relations are steeper than observed for MW clouds. \cite{Tasker:2009aa} argued that low numerical resolution was the culprit, and while resolution likely plays a role, these results agree with our findings in \S\ref{sect:results} that stellar feedback is a necessary ingredient for reproducing observed properties of GMCs. In this picture, turbulence is injected at $\sim \kpc$ scale by gravitational processes (instabilities, shear), and cascades down to parsec scales, where star formation and stellar feedback rapidly lead to GMC dispersal and further turbulence driving \citep[][]{Padoan:2016aa}. 

As shown in G17, feedback redistributes gas from small ($\lesssim100\kpc$) to large ($\gtrsim\kpc$) scales by heating and accelerating gas on small scales, driving galactic winds (see Fig.~1 of G17), thus recycling gas back to large galactic scales \citep[see also][]{Semenov:2017aa}.  In the context of our study, once stars form feedback heats and removes gas from GMCs and their surroundings, limiting further growth. In essence, feedback is a crucial component in establishing sustained gas recycling. To fully determine how feedback changes the Larson relations would require tracking the GMCs throughout their lifetime which is beyond the scope of this work, but will be addressed in future papers in this series (Grisdale et al., in prep.).

It is evident that the way in which stellar feedback is implemented matters. \cite{Fujimoto:2016aa} studied GMC properties using global disc simulations with a stellar feedback based on purely thermal injection of SNe energy. Such models are known to suffer from overcooling, leading to inefficient turbulence driving and regulation of star formation in cold gas \citep[see e.g.][]{Agertz:2013aa,Hopkins:2014aa}. Indeed, they found a rather weak effect of feedback on GMC properties compared to their simulation without \footnote{We note that their thermal feedback model reduced the fraction of massive clouds found in the bar region, with little effect throughout the disc.}. 

Our adopted feedback prescription is much more efficient; it succeeds to both reproduce observed GMC properties as shown in previous sections, and at the same time predicts the existence of powerful galactic outflows in star forming high redshift galaxies, leading to $z=0$ disc galaxies compatible with observations \citep[][]{Agertz:2015ab,Agertz:2016aa}. The need for efficient feedback in predicting GMC properties was also demonstrated using full galactic models by \cite{Hopkins:2012aa} using momentum feedback, and more recently in the M33 models by \cite{Ward:2016aa}\footnote{also basing their feedback model on \cite{Agertz:2013aa}}

While we have shown that efficient feedback is key in reproducing GMC properties in galactic models, the role of gravity is still not yet understood. Gravitational instabilities coupled with galactic shear are capable of reproducing the observed level of turbulence in {\small HI} \citep[e.g.][]{Agertz:2009aa,Bournaud:2010aa,Krumholz:2016aa}, its role in establishing GMC properties is a less clear but hotly debated subject. \cite{Padoan:2016aa} demonstrated that purely SNe driven turbulence, in a patch of the ISM, leads to a population of GMCs and scaling relations in agreement with observations. Similar results were found for pure solenoidal large scale turbulence driving by \cite{Saury:2014aa}. In contrast, \cite{Ibanez-Mejia:2016aa} argued, using simulations of stratified ISM patches, that self-gravity was an essential component for establishing GMC scaling relations. Our models do not resolve this issue, but makes it clear that gravity alone does not allow for a sustained coupling of small and large scales over galactic dynamical times (100s of Myrs).

Finally, we note that it may seem surprising that our galactic simulations, reaching a spatial resolution of $\sim 4.6\pc$, can account for GMC properties in clouds that typically are resolved with only $\sim 100$ cells \citep[see also work based on SPH by][]{Baba:2017aa}. If resolving the internal density and velocity structure of clouds was essential, much higher spatial resolution is likely needed \citep[e.g.][]{Padoan:2016aa,Seifried:2017aa}. We speculate that this illustrates that as long as a realistic turbulent coupling can be established between large (kpc) and small scales\footnote{where small scale physics are captured robustly by  subgrid models \emph{below} the scale of where the GMC analysis takes place} \citep[as demonstrated for our simulations in][]{Grisdale:2016aa}, \emph{global} properties of GMCs will be captured. This notion is compatible with analytical work by e.g. \cite{Hopkins:2012ab}.

\section{Conclusions}
\label{con}

In this work we explore the origins of physical properties of Giant Molecular Clouds (GMCs) using hydrodynamical simulations of Milky Way-like galaxies, with the goal of understanding the role of stellar feedback. To this end we analyse two simulations, one with stellar feedback and one without \citep[see also][]{Grisdale:2016aa}. GMCs are identified in projection (2D), as well as in 3D, using clump finders, and for each cloud we measure its physical properties, e.g. mass ($M$), surface density ($\Sigma$), radius ($R$) and velocity dispersion ($\sigma$). We investigate how stellar feedback affects GMC scaling relations (`Larson's relations') and discuss the turbulent coupling between large and small galactic scales that lead to observed GMC properties.
Our key results are summarised below:

\begin{enumerate}
	
	\item The galaxy simulations with stellar feedback produce a population of clouds that, when detected and analysed in projection, are in excellent agreement with observed Milky Way GMCs \citep{Heyer:2009aa}.

	\item Feedback regulates the properties of GMCs, limiting the maximum mass, radius, surface density, velocity dispersion and level of boundness (as quantified by the virial parameter $\ap$), in a cloud population.
		
	\item We measure Larson's scaling relations, $\sigma\propto R^{a},\,M\propto R^{b}$ and $\sigma\propto(R\Sigma)^{c}$, for our cloud populations and find that these relations depend strongly on the presence of stellar feedback. With stellar feedback, GMC scaling relations in the simulation are in strong agreement with relations for Milky Way clouds \citep{Heyer:2009aa}, with $(a,b,c)\sim(0.56,2.3,0.41)$. The simulation without feedback predicts significantly steeper scaling relations which is incompatible with observations.
			
	\item When detected and analysed using a 3D clump finder we find different cloud properties. Clouds tend to be more massive (an increase by a factor of $\sim 2$) and therefore feature higher surface densities. This causes a steepening in GMC scaling relations, where the largest effect is found for the $M$-$R$ relation.
			
	\item We analyse in detail how the virial parameter can affect derived GMC scaling relations. In particular, we highlight how the underlying $\ap$ distribution can affect measured slopes of GMC scaling relations. We find that our simulation with stellar feedback produces  a distribution of $\ap$ compatible with observed Milky Way clouds, with massive clouds being more bound ($\ap\propto M^d$ with $d<0$), with an opposite trend found in the simulation without feedback.

	\item Clouds with nearly identical global properties can exist in different evolutionary stages. We show that at all times, the feedback driven ISM results in a near constant $\sim 1/4$ of GMCs in a state of collapse, $\sim 1/5$ compressed by external ISM pressure, and the rest being in a bound or unbound state, with little influence from external pressure. 

\end{enumerate}

To conclude, we have demonstrated that stellar feedback acts to regulate physical properties of GMCs. To accurately model galaxies from scales of tens of parsecs up to kiloparsecs requires both a realistic accounting, and modelling, of energy and momentum input from massive stars into the ISM, as well as large scale turbulence driving by gravity and shear.

The work presented here is the first step towards an in depth exploration of the impact of feedback on the formation, evolution and destruction of GMCs. In future work we will explore the origins of observed GMC star formation efficiencies. In a second work we will quantify the evolution of GMC properties, and how individual clouds 'move' in the $M$--$R$, $\sigma$--$R$ and $\sigma^{2}/R$--$\Sigma$ planes over their life times.

\section*{acknowledgments}
We thank the anonymous referee for their valuable comments. KG thanks the University of Surrey for his studentship. OA acknowledges support from the Swedish Research Council (grant 2014- 5791). KG and FR acknowledges support from the European Research Council through grant ERC-StG-335936. OA and FR also acknowledges support from the Knut and Alice Wallenberg Foundation. 

\bibliographystyle{mn3e}
\bibliography{ref}

\appendix

\section{Resolution Effects}
\label{App:resolution}
\begin{figure*}
	\begin{center}
		\includegraphics[width=0.85\textwidth]{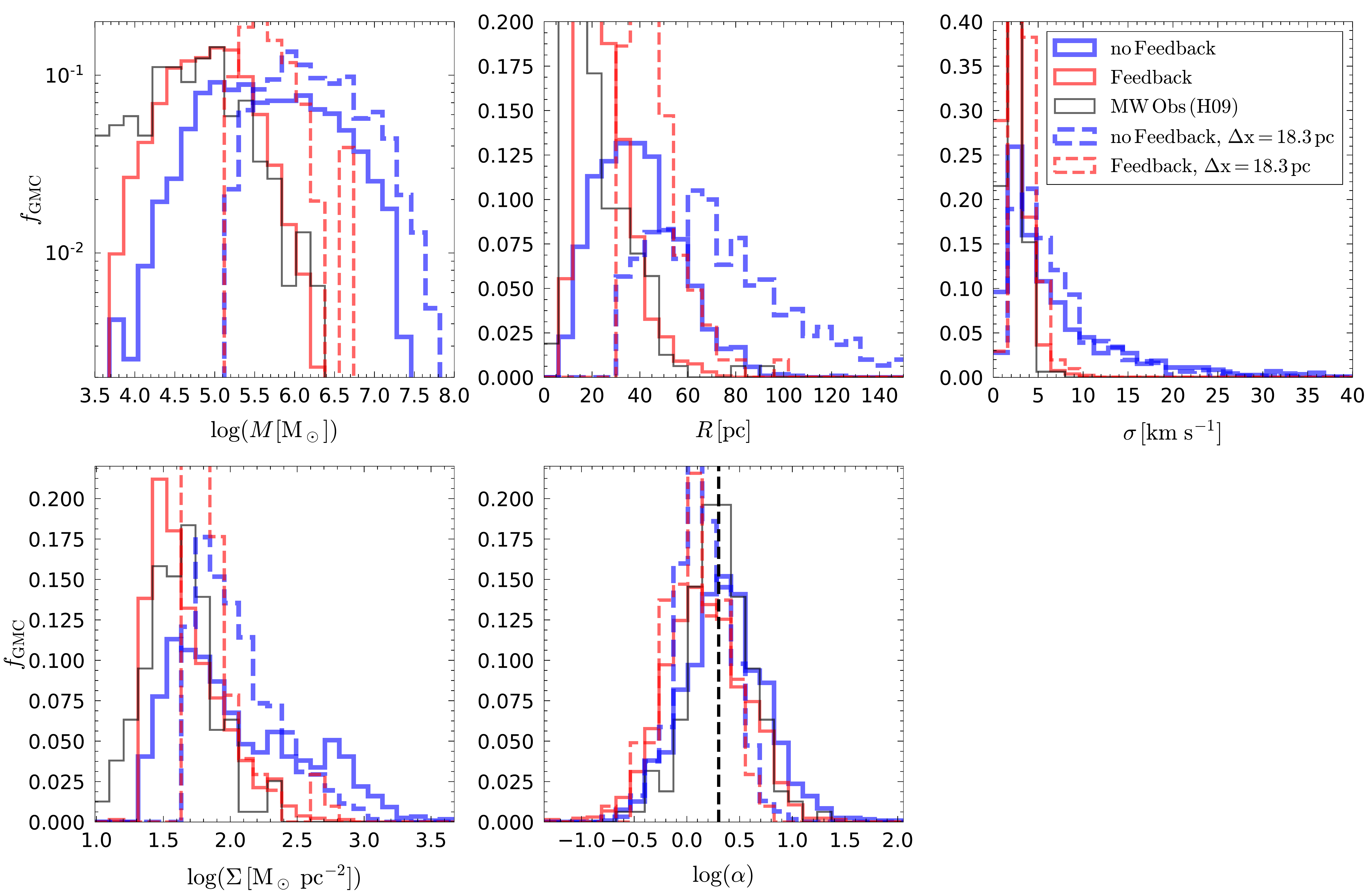}
		\caption{Comparison of the normalised distributions of the properties (mass, radius, velocity dispersion, surface density, virial parameter) of GMCs detected with the 2D method at two resolutions. The red, blue and black lines show data from the simulations with feedback, without feedback and data from H09 respectively. Data from from our fiducial simulation (i.e. $\Delta x\sim4.6\pc$) are shown with solid lines, while those from the low resolution run ($\Delta x\sim18.3\pc$) are shown with dashed lines. The dashed vertical line in the middle-bottom panel shows $\ap=2$, i.e the border between bound and unbound. 		
		}
		\label{fig:2D_res}
	\end{center}
\end{figure*}

\begin{figure*}
	\begin{center}
		\includegraphics[width=0.85\textwidth]{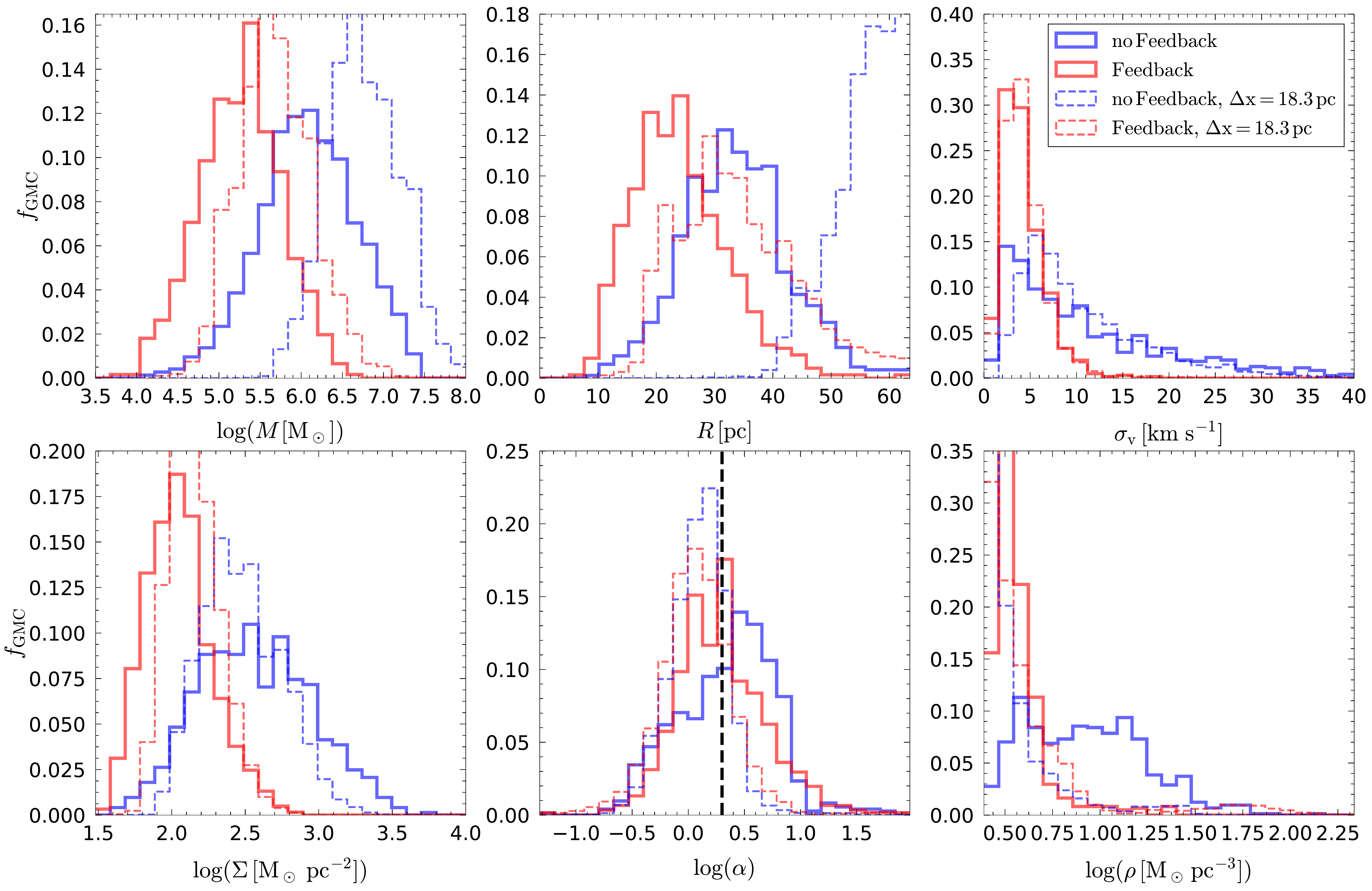}
		\caption{Same as Fig.~\ref{fig:2D_res}, but for GMC detected using the 3D clump finder. 
		}
		\label{fig:3D_res}
	\end{center}
\end{figure*}

To explore the role of numerical resolution on the results described in \S\ref{sect:results}, a second series of simulations were run using the same initial conditions, code, feedback prescription, star formation recipe, refinement criteria and clump finder settings, but with a maximum resolution of $\Delta x\sim18.3\pc$, i.e. a $4\times$ coarser  than in the fiducial simulation. Fig.~\ref{fig:2D_res} and \ref{fig:3D_res} compares the high and low resolution results from the 2D and 3D clump finding methods (see \S\ref{method:2dclumps} and \ref{method:3dclumps}) at $t=175\Myr$.

When identified in 2D (Fig.~\ref{fig:2D_res}) the change in spatial resolution changes the range and distribution of the masses, radii and surface densities of GMCs, however the velocity dispersions and virial parameters remain relatively unchanged. The change in GMC sizes is expected when changing resolution, i.e. clouds smaller than the resolution of the simulation will merge into one unresolved structure, as shown by the sharp truncation GMC radii at $\sim30\pc$. Furthermore, the minimum number of cells per cloud allowed by the clump finder (here 9), also strongly affects the size distribution. As the smallest clouds in the low resolution run are an amalgamation of one or more smaller clouds in the fiducial resolution run they naturally also have larger masses. This truncates the mass distribution at a few $10^{5}\Msol$. 

The 3D clump finding results (i.e. Fig.~\ref{fig:3D_res}), also shows some differences between the two resolutions. However in this case the differences is primarily limited to the mass and radius distributions. In both cases the distribution is shifted to larger values at lower resolution, but the shape of the distribution is largely unchanged. The result of both distributions being shifted to larger values results is a mostly unchanged distribution of surface densities.  As found by the 2D clump finder, the velocity dispersion of the clouds is relatively unchanged and therefore the low resolution simulations gives GMCs with similar virial parameter distributions to their high resolution counterparts.

\end{document}